\documentstyle[12pt]{article}

\newcommand {\abc}[1]{{\mathop {\mbox {$#1$}}\limits_{\sim{}}}}

\begin{document}

\title{On classical description of radiation from neutral fermion with
anomalous magnetic moment}

\author{A.E. Lobanov
and O.S. Pavlova
\thanks{E-mail: $OSPavlova@aport.ru$}}
\maketitle

\begin{center}
{\em Moscow State University,\\
Department of Theoretical physics.\\
$117234$, Moscow, Russia}
\end{center}

\begin{abstract}
Electromagnetic radiation from an uncharged spin 1/2 particle with
an anomalous
 magnetic moment moving in the classical electromagnetic external field
originates from quantum spin-flip transitions. Although this process has a
 purely quantum nature, it was observed for certain particular external
field configurations that, when quantum recoil is neglected,
 the radiation power corresponds
to the classical radiation from an evolving  magnetic dipole. We argue that this
 correspondence has a more general validity in the case of an
{\it unpolarized} particle
 and derive a general formula for radiation in terms of the external
 field strength and its derivative. A classical dynamics
of the spin is described by the Bargmann-Michel-Telegdi equation.
\end{abstract}
\vskip 10pt
{PACs: 03. 65. Sq}\vskip15pt

\sloppy

Electromagnetic radiation from an uncharged spin 1/2 particle possessing an
 anomalous magnetic moment moving in an external classical electromagnetic
 field was studied in a number of papers [1-8]. The usual approach consists in a
computation of the radiative transitions within the Furry's picture of quantum
 electrodynamics (QED). This appeals to the knowledge of exact solutions
 of the Dirac equation with an anomalous magnetic moment in the external
 electromagnetic field.
 Such solutions are known for various particular configurations of
 the electromagnetic field: homogeneous fields [1,2], plane-waves [3-6], and
some other [7,8], for them an explicit computation of radiative transitions
was performed. Radiative transitions in the case of the uncharged
 particles
correspond to the spin-flip amplitudes and thus have an essentially quantum nature.
Meanwhile it was observed that under certain conditions the radiation
process may be described
in purely classical terms using the  Bargmann-Michel-Telegdi (BMT) spin
evolution equation
 [9, 10]. In such a pseudoclassical treatment
the radiation power is given by the well-known formula for radiation from
a magnetic moment $\mu^\nu$ [11]:
\begin{equation}
  \frac {dI}{d{\cal O}}=
 -\frac {1}{4\pi(lu)^5u^0}\{{{\ddot \mu}^\nu}
{{\ddot \mu}_\nu}(ul)^2+(l^\nu {\ddot \mu}_\nu)^2\}
\end{equation}
where $(ul)=u^\nu l_\nu$, $u^\nu$ is the 4-velocity of a particle,
$l^\nu=\{ 1,{\bf l}\},$ ${\bf l}$ is the unit vector in the direction of
 an emitted wave;
a dot
denotes the differentiation with respect to the proper
time $\tau$. We use the units $\hbar =c=1 $.

The radiation power calculated within the framework of QED was found
to correspond to
the result obtained from the equation (1) under the following conditions.
 The particle motion has to be quasiclassical, in a sense that:

 i) the binding energy due
the magnetic moment in the rest frame is much smaller than the particle
 mass (here we use gaussian units):
\begin{equation}
{H_0}\ll {H_{cr}},\hspace{0.5cm}  H_{cr}={m^2c^3}/{e\hbar},
\end{equation}
($H_0$ is the magnetic field strength in the rest frame),

ii) the external field varies slowly at the distances of the order of a
Compton's length, which amounts to conditions:
\begin{equation}
 {\hbar\dot {H}_0}/{mc^2H_0}\ll 1,\hspace{0.9cm}
({H_0}/{H_{cr}})({\omega}/{\omega_c})\ll 1,
 \end{equation}
  (here $\omega$ is
the characteristic frequency change of the external field,\, and
$\omega_c={eH}/{mc}$ is the cyclotron frequency).

If one is interested in the radiation from an unpolarized particle, an
additional requirement to be imposed consists in averaging over
the initial spin states and summing over the final polarizations.
It can be expected that the averaging of the
quantum transition amplitudes should correspond to the averaging over
the initial orientation of the magnetic dipole moment within the classical
 picture.

Although no general derivation of such a pseudoclassical description from
QED is available for the case of an arbitrary external field, it seems
very plausible that under the above assumptions the result should follow
indeed
from the classical formula (1). Our conjecture is that one has to replace
the magnetic moment by the quantity
\begin{equation}
\mu^\nu=\mu_0 S^\nu,
\end{equation}
where $S^\nu$ is expectation value of the spin vector, and perform an
averaging over the polarizations at $\tau=\tau_0$:
$$
  \frac {dI}{d{\cal O}}=
 -\frac {\mu^2_0}{4\pi(lu)^5u^0}<\{{{\ddot S}^\mu}
{{\ddot S}_\mu}(ul)^2+(l^\mu {\ddot S}_\mu)^2\}>=
$$
\begin{equation}
=-\frac {\mu^2_0}{4\pi(lu)^5u^0}<e^{\mu\nu\rho\lambda}
{\ddot S}_{\nu} u_{\rho} l_{\lambda} e_{\mu{\nu^\prime}{\rho^\prime}
{\lambda^\prime}}{\ddot S}^{\nu^\prime} u^{\rho^\prime} l^{\lambda^\prime}>,
\end{equation}

An evolution of the classical spin is described by the BMT equation [12]:
\begin{equation}
\dot S^\nu=2\mu_0\{F^{\nu\alpha}S_\alpha -u^\nu (u_\alpha
F^{\alpha\beta}S_\beta)\},
\end{equation}
whose solution has to be constrained  by the conditions $S^\nu S_\nu=-1$,
\mbox{ $S^\nu u_\nu=0$}. The validity of this equation is ensured by the
conditions (2), (3)  [13, 14].
Our main goal is to show that when the averaging over polarization states
is performed the resulting expression for the radiation power will depend
only on the external field intensity and thus will be valid in the case
 of arbitrary external field subject to conditions (2),(3). We check
 that it is true indeed for the particular fields studied
before.

An important point in the derivation is that the neutral particle
moves in the external field with a constant velocity.
  Of course, the true quantum description of radiation demands that
the quantum recoil in the
 photon emission process should be taken into account. But when conditions
 (2), (3) are satisfied, energy of emitted photon is small, therefore we
can neglect the recoil, i.e. the change of
 particle velocity. This assumption was made in the formula (5).
In the BMT equation one can pass to the
rest frame of the particle where the required averaging over
 polarizations can be easily performed.

It is more convenient to rewrite the BMT equation in term of the
$SL(2,\bf{C})$ spin-tensors.
 Instead of the 4-vector $S^\nu$ we will use  matrices
 ${\abc S}=S^0 \sigma_0+\mbox{\boldmath
$S\sigma$}$ and $\tilde S=S^0 \sigma_0-\mbox{\boldmath
$S\sigma$}$, where $\sigma_\nu$ are the Pauli matrices
(we follow a notation of [15]).

Then the  evolution of spin is given by the following matrix operator [16]:
\begin{equation}
{\abc S}(\tau)={\abc L}{\abc R}{\abc L}^{-1}{\abc {S_0}}({\abc
L}^{-1})^{+}{\abc R}^{+}{\abc L}^{+}.
\end{equation}
Here $S_0\equiv S(\tau_0)$ is the polarization at the initial
moment of time $\tau_0$,  ${\abc L}^{-1}$ is a transformation to
the particle rest frame
$$
{\abc L}^{-1}=\frac{1+\tilde u}{\sqrt{2(1+u^0)}},\hspace{1cm}
{\abc L}=\frac{1+{\abc u}}{\sqrt{2(1+u^0)}};
$$
and ${\abc R}$ is a rotation operator satisfying the equation
\begin{equation}
\dot {\abc R}=i\mu_0 {\abc {H_0}} {\abc R},
\end{equation}
where
\begin{equation}
{\bf H}_0(x)\equiv{\bf H}_0=u^0{\bf H}-{\bf u}\times{\bf E}-\frac
{{\bf u} ({\bf u}{\bf H})}{1+u^0}\hspace{ 0.25cm}-
\end{equation}
is the magnetic field in the rest frame at the point of the particle
location.
In this notation the formula (5) can be presented in the form:
 \begin{equation}
\frac {dI}{d{\cal O}}=
 -\frac {\mu^2_0}{8\pi(lu)^5 u^0}<Sp\{{\ddot \abc S}{\ddot {\tilde S}}{(ul)^2}+
{\frac {1}{4}} [{\abc l}{\ddot {\tilde S}}{\abc l}{\ddot {\tilde S}}+
{\ddot \abc S}{\tilde l}
{\ddot \abc S}{\tilde l}]\}>.
\end{equation}
After a simple algebra with account for the Eq.(7) we obtain
 for the radiation
power of unpolarized particle:
\begin{equation}
\frac {dI}{d{\cal O}}=
 -\frac {\mu^2_0}{16\pi(lu)^3 u^0} Sp\sum_{i}\{\ddot{(\sigma_R)_i}
\ddot{(\sigma_R)_i}-\ddot{(\sigma_R)_i}
(\mbox{\boldmath $\sigma n$})
\ddot{(\sigma_R)_i}(\mbox{\boldmath $\sigma n$})\}.
\end{equation}
where
$$
(\sigma_R)_i={\abc R}\sigma_i{\abc R}^{+}, \hspace{1cm}
 {\bf n}={\frac {1}{(ul)}}\{{\bf l}-{\frac {1+(ul)}{1+u^0}}{\bf u}\} -
$$
is the unit vector in the direction of radiation in the rest
 frame of the particle. In (11) the summation over $i$ corresponds to
an averaging over the initial
spin states and  summation over the final spin states.

Using Eqs. (8) and (11) we arrive at the final result
\begin{equation}
\frac{dI}{d\cal O}=\frac{\mu^4_0}{\pi u^0 (ul)^3}\{4{\mu_0}
\{ \mu_0 {\bf H}^4_0+{\mu_0}{\bf H}^2_0({\bf H}_0 \bf n)^2+
({\bf H}_0 \bf n)[{\dot {\bf H}_0} {\bf H}_0{\bf n}]\}+
{\dot{\bf H}_0}^2+({\dot{\bf H}_0}{\bf n})^2\}.
\end{equation}
Therefore the angle distribution of radiated power is expressed entirely
 in terms of the magnetic field strength in
the rest frame of the particle and its derivative
with respect to the proper time.

 Formula (12) can be rewritten in a covariant form:
\begin{equation}
\begin{array}{rl}
{}&{}\\
\frac{dI}{d\cal O}=&\frac{\mu^2_0}{\pi u^0 (ul)^5}
\{\lbrack 4({\mu^2_0}{u_\rho}{H^{\rho\lambda}}{H_{\lambda\sigma}}{u^\sigma})^2+
({\mu^2_0}{u_\rho}{{\dot H}^{\rho\lambda}}{{\dot H}_{\lambda\sigma}}{u^\sigma}) \rbrack (ul)^2\\
{}&{}\\
{}&+4({\mu^2_0}{u_\rho}{H^{\rho\lambda}}{H_{\lambda\sigma}}{u^\sigma})
({\mu_0}{u_\rho}{H^{\rho\lambda}}l_{\lambda})^2+
({\mu_0}{u_\rho}{{\dot H}^{\rho\lambda}}l_{\lambda})^2\\
{}&{}\\
{}&+4\mu^2_0({\mu_0}{u_{\rho}}{H^{\rho\lambda}}l_{\lambda})e^{\mu\nu\rho\lambda}
u_{\mu}{\dot H}_{\nu\sigma}{u^{\sigma}}{H_{\rho\delta}}{u^{\delta}}
{l_{\lambda}}\},\\
{}&{}\\
\end{array}
\end{equation}
where $H^{\nu\alpha}=-\frac 12 e^{\nu\alpha\beta\gamma}F_{\beta\gamma}$
is the tensor dual to electromagnetic field tensor.

Integrating over angles, we obtain the formula
 for the total radiation power:
\begin{equation}
I=\frac{16}{3}\mu^2_0\{ 4(\mu_0 {\bf H}_0)^4 + (\mu_0 {\bf \dot H}_0)^2 \},
\end{equation}
 which has the following covariant counterpart:
\begin{equation}
I=\frac{16}{3}\mu^2_0 \{4({\mu^2_0}{u_\rho}{H^{\rho\lambda}}
{H_{\lambda\sigma}}{u^\sigma})^2+
{\mu^2_0}{u_\rho}{{\dot H}^{\rho\lambda}}{{\dot H}_{\lambda\sigma}}
{u^\sigma} \}.
\end{equation}

We verified that, substituting into formulas (12), (14), the fields which were
earlier used for calculations both by quantum and classical methods, we
obtain the full agreement with the results obtained when we neglect
the recoil,
i.e., terms of the higher order in the Planck constant.

Now let us discuss some possible application for a neutron. An important
features of the Eqs. (12) and (14) is the presence of the second term,
whose magnitude is defined by the field strength and its derivative.
Therefore the radiation
power
is increased substantially when the boundary of field region which is
crossed
 by a particle is very sharp. This phenomenon similar to
 transition radiation can be important in astrophysical conditions.
 Possibly this effect can be observed under the laboratory
 conditions when fast neutrons pass through the ferromagnetic.

It is interesting to compare in the order of magnitude the radiation power
$I$ from a neutral particle (neutron) and the classical radiation power
$I_0$ from a charged particle (see, for example, [11]) with the same mass
and the energy (proton). We find
$$
\frac{I}{I_0}\sim max\left\{\left(\frac{H_0}{H_{cr}}\right)^2,
\left(\frac{u^0\hbar\omega}{mc^2}\right)^2\right\}
$$
Thus, the charge radiation power and the radiation power from a magnetic
moment become comparable in very strong fields, or in the fields of high
frequency.


\vspace{1cm}

\noindent
{\bf Acknowledgments.}

\vspace{0.7cm}

Authors are very grateful to Prof.~D.~V.~Gal'tsov for continuous attention
to their work, Prof.~A.~V.~Borisov and
Prof.~V.~Ch.~Zhukovsky for fruitful discussions.


\end{document}